\documentclass[11pt]{article}

\def\la{\leftarrow}
\def\ra{\rightarrow}
\def\ex{\exists}
\def\fa{\forall}
\def\sbd{\vspace{8pt}\noindent\bf}

\newbox\tempa
\newbox\tempb
\newdimen\tempc
\def\mud#1{\hfil $\displaystyle{\mathstrut #1}$\hfil}
\def\rig#1{\hfil $\displaystyle{#1}$}
\def\irulehelp#1#2#3{\setbox\tempa=\hbox{$\displaystyle{\mathstrut #2}$}%
		        \setbox\tempb=\vbox{\halign{##\cr
	\mud{#1}\cr
	\noalign{\vskip\the\lineskip}%
	\noalign{\hrule height 0pt}%
	\rig{\vbox to 0pt{\vss\hbox to 0pt{${\; #3}$\hss}\vss}}\cr
	\noalign{\hrule}%
	\noalign{\vskip\the\lineskip}%
	\mud{\copy\tempa}\cr}}%
		      \tempc=\wd\tempb
		      \advance\tempc by \wd\tempa
		      \divide\tempc by 2 }
\def\irule#1#2#3{{\irulehelp{#1}{#2}{#3}%
		     \hbox to \wd\tempa{\hss \box\tempb \hss}}}

\begin{document}
\begin{center}
{\Large{\bf Automatic Proof Checking and Proof Construction by Tactics}}
\\[10pt]
{\bf Notes for the Workshop on Meta-Variables, September 1991.}\\[10pt]
{\bf Gilles Dowek}\\[10pt]
{\bf INRIA}\def\thefootnote{\fnsymbol{footnote}}\footnote[1]{B.P. 105, 78153
Le Chesnay CEDEX, France. dowek@margaux.inria.fr}\footnote[2]{This research
was partly supported by ESPRIT Basic Research Action ``Logical Frameworks''.}
\def\thefootnote{\arabic{footnote}}\\
\end{center}

In this note we compare two kinds of systems that verify the correctness of 
mathematical developments: {\it proof checking} and {\it proof construction by
tactics} and we propose to merge them in a single system.
We consider mathematics formalized in the Calculus of Constructions 
\cite{Coquand85} \cite{CoqHue88} but the ideas given here are not specific to
this formalism.
As an example of a proof checking system we consider the {\it constructive 
engine} of the system Coq \cite{Huet89} and as an example of a proof 
construction
by tactics system we consider the {\it proof assistant} of the system Coq 
\cite{Coq} designed in the spirit of the system LCF \cite{GoMiWa}.

\section{Motivations}

\subsection{The Constructive Engine and the Proof Assistant}

The Constructive Engine and the Proof Assistant are both systems that aim to 
give a tool to a mathematician to write a theorem and its proof and check the 
correctness of the proof.

With the {\it constructive engine} the user gives the proof of the theorem 
as a $\lambda$-term. The answer of the machine is a binary information:
$right$ or $wrong$ according to the correctness of the proof.
For instance, the following developments are correct.

Example 1:

\begin{verbatim}
Theorem I.
Statement (P:Prop)(P -> P).
Proof [P:Prop][x:P]x.
\end{verbatim}

Example 2:

\begin{verbatim}
Parameter T:Prop.

Parameter R:T -> T -> Prop.

Parameter Eq:T -> T -> Prop.

Axiom Antisym:(x:T)(y:T)(R x y) -> (R y x) -> (Eq x y).

Parameter a:T.

Parameter b:T.

Axiom ax1:(R a b).

Axiom ax2:(R b a).

Theorem th.
Statement (Eq a b).
Proof (Antisym a b ax1 ax2).
\end{verbatim}
With the {\it proof assistant} the machine is more active, the theorem to be 
proven is taken as a {\it goal}. The set of goals is transformed by 
{\it tactics}, a tactic is a function that maps a set
$G$ of goal to a set $G'$ of goals such that proof of the goals of $G$ can be 
constructed from proofs of the goals of $G'$.
Proving a proposition consists in applying tactics until we get an empty set of
goals. The same examples are written as follows.

Example 1:

\begin{verbatim}
Goal (P:Prop)(P -> P).

Intro.
                    1 subgoal
                      P->P
                      ============================
                        P : Prop

Intro.
                    1 subgoal
                      P
                      ============================
                        H : P
                        P : Prop

Apply H.
                    Goal proved!
\end{verbatim}

Example 2 (we keep the same declarations of parameters and axioms):

\begin{verbatim}
Goal (Eq a b).

Apply Antisym.
                      (R a b)
                    subgoal 2 is:
                      (R b a)

Apply ax1.
                      (R b a)

Apply ax2.
                    Goal proved!

\end{verbatim}
Let us consider the first example.
We first have to prove the proposition $(P:Prop)(P \ra P)$, 
the tactic $Intro$ transforms this goal into the goal 
$P \ra P$ in a context where $P$ is a
proposition then the same tactics transforms the goal in $P$ in a context
where $P$ is a proposition and $H$ a proof of $P$. Then the 
tactic $Apply~H$ finds the proof $H$ for the goal $P$ and 
the set of remaining subgoal is empty, so the proof is over.

Then from this sequence of tactics we can build the proof 
$[P:Prop][H:P]H$.
This proof is built in three steps: when we first apply the tactic 
$Intro$ we build the proof $[P:Prop]?y$ where $?y$ is a
meta-variable
denoting the forthcoming proof of the current goal $P \ra P$.
When we apply another time the tactic intro, we build the proof
$[P:Prop][H:P]?z$ where $?z$ is a meta-variable
denoting the forthcoming proof of the current goal $P$.
When we apply the tactic $Apply~H$ we get the proof
$[P:Prop][H:P]H$.

When the goal is proved, the theorem can be added to the context of the 
engine with the command:
\begin{verbatim}
Save I.
\end{verbatim}
The symbols $?y$, $?z$ are called meta-variables, they denote 
proof-terms to be constructed. When such a proof term is constructed it is 
substituted to the meta-variable, remark that this substitution may capture 
variables for instance when we substitute $H$ to $?z$ in 
the term $[P:Prop][H:P]?z$ we get the term $[P:Prop][H:P]H$.

In a proof assistant, the tactic $Intro$ and the tactic $Apply$
can be distinguished as basic tactics. The 
tactic $Intro$ transforms the goal $(x:A)B$ in the goal $B$
adding $x:A$ to the context. The tactic $Apply$
uses a known proposition $(x_{1}:P_{1}) ... (x_{n}:P_{n})P$ and a goal $Q$,
unifies $P$ and $Q$ (this unification binds some of the $x_{i}$'s),
and gives back as subgoals the $P_{i}$'s such that $x_{i}$ is not bound by 
unification.
For instance in the second example we want to prove
$(Eq~a~b)$ we unify $(Eq~x~y)$ and $(Eq~a~b)$ and we generate the subgoals
$(R~a~b)$ and $(R~b~a)$ the proof associated to this step is
$(Antisym~a~b~?v~?w)$ where $?v$ and $?w$ are proofs of $(R~a~b)$ and 
$(R~b~a)$ to be found.

\subsection{Merging Proof Checking and Proof Construction by Tactics}

Most of mathematics verifying systems as for instance the system Coq
\cite{Coq} can be used in two modes: proof checking and proof construction by
tactics. We design a system with a single mode and such that:

$\bullet$ The declaration of the statement of a theorem:
\begin{verbatim}
Theorem I.
Statement (P:Prop)(P -> P).
\end{verbatim}
and of a goal:
\begin{verbatim}
Goal (P:Prop)(P -> P).
\end{verbatim}
are done in the same way.

$\bullet$ The user can work on a proof, go to another one come back
to the first, etc.

$\bullet$ The user can work by tactics (top-down) on a goal, decide to stop
to prove a lemma (bottom-up) and go back to its proof.

$\bullet$ Since unification is undecidable we do not want the
tactic $Apply$ to be restricted by a subcase of unification (as first order 
unification)
but we want to let the user guide the machine in the unification tree as he 
does in the proof tree.

\section{Meta-variables are Variables: An Engine}

\subsection{Meta-variables are Variables}

When we have considered incomplete proofs as $(Antisym~a~b~?v~?w)$ we have 
added to the formalism of terms symbols $?v$, $?w$ that denotes some
terms. In the same way, in elementary algebra when we can write an equation
as $x^{2} + 6 = 5 x$ we add to the formalism of integers a symbol $x$ that 
denotes an integer. The expression $x^{2} + 6$ is an incomplete integer. 
The symbol $x$ is called a variable.

In contrast with the formalism of numbers, the $\lambda$-calculus already has 
a notion of variable. So adding a new notion of meta-variables is 
not useful and we can write this incomplete proof $(Antisym~a~b~v~w)$.
As far as term formation and type checking are concerned the variables $v$ and 
$w$ are not different from $a$ and $b$, in particular $v$ and $w$ {\it have to 
be declared} in a context $\Gamma$ where $(Antisym~a~b~v~w)$ is well-formed.

The variables $v$ and $w$ are distinguished from $a$ and $b$ only for 
substitution since $v$ and $w$ can be instanciated and $a$ and $b$ cannot.
This information is expressed in adding a quantifier to each variable in
the context, the universal quantifier $\fa$ is added to the declaration of 
variables that cannot be instanciated (as $a$ and $b$) and the existential 
quantifier $\ex$  is added to the declaration of the variables that can be 
instanciated (as $v$ and $w$).

\subsection{Constraints}

When we apply a substitution to a variable $x$
the term substituted to $x$ must have the same type as $x$.
Let us consider the context:
$$\Gamma = [\fa A:Prop; \ex X:Prop; \ex y:X \ra A]$$
a priori we cannot substitute the term $[x:X]x$ to the variable $y$
since the term $[x:X]x$ has type $X \ra X$ and $y$ has type $X \ra A$.
But if we apply first the substitution $X \la A$, the substitution $y \la [x:A]x$
is allowed.

So before applying the substitution  $y \la [x:X]x$ we should unify 
the type of $y$ and the type of $[x:X]x$.
Since unification is in general a difficult problem, we do not want to have 
to perform a
unification step before each substitution, so we allow the substitution
$y \la [x:X]x$ but we will keep a constraint $X \ra X = X \ra A$ to 
remind that this equation has to be solved even if we not want to solve it now.

In the same way, when we have an existential variable (a goal) 
$g:(Eq~a~b)$ we may 
introduce new existential variables $h_{1}:T$, $h_{2}:T$, 
$h_{3}:(R~h_{1}~h_{2})$, $h_{4}:(R~h_{2}~h_{1})$ and 
instanciate the variable $g$ the term
$(Antisym~h_{1}~h_{2}~h_{3}~h_{4})$.
The generated constraint is $(Eq~a~b) = (Eq~h_{1}~h_{2})$. We do not have to 
solve this equation to perform the substitution, but this equation will 
restrict the substitutions to be performed for $h_{1}$ and $h_{2}$ in the 
future.

If, as in this case, the constraint is a first order unification problem 
it can be solved and the substitution $h_{1} \la a, h_{2} \la b$ can 
be performed. But if the constraint is a higher order unification problem then
the user can propose substitutions for the variables occurring in the equation
and guide the machine in the unification tree as he does in the proof 
tree\footnote{This
view of unification as constrained resolution leads in \cite{complete}
to a complete proof synthesis method where unification and resolution are
merged in a single algorithm.}.

\subsection{Constrained Quantified Contexts}

{\sbd Definition:} Constrained Quantified Contexts 

A {\it quantified declaration} is a triple $<Q,x,T>$ (written $Qx:T$) 
where $Q$ is a quantifier ($\fa$ or $\ex$), $x$ a symbol and $T$ is a
term. 
A {\it constant definition} is a triple $<x,t,T>$ (written $x := t:T$)
where $x$ is a symbol and $t$ and $T$ terms.
A {\it constraint} is a pair of terms $<a,b>$ (written $a = b$).
A {\it constrained quantified context} is a list
$\Gamma = [e_{1}; ...; e_{n}]$ such that $e_{i}$ is either 
a quantified declaration a constant definition or a constraint. These 
constrained quantified contexts are generalization of Miller's 
{\it mixed prefixes} \cite{Miller91a} which are lists of quantified 
declarations.

Non constrained, non quantified contexts are identified with constrained 
quantified contexts with only universal variables and constants.

{\sbd Definition:} Equivalence Modulo Constraints

Let $\Gamma$ be a constrained quantified context, we
define the relation between terms $\equiv_{\Gamma}$ as the smallest 
equivalence relation compatible with terms structure such that:
\begin{itemize}
\item{if $t \equiv t'$ then $t \equiv_{\Gamma} t'$,}
\item{if $(a = b) \in \Gamma$ then $a \equiv_{\Gamma} b$.}
\end{itemize}

{\sbd Definition:} Typing Rules

First we modify the rules of the Calculus of Constructions to deal with the 
new syntax:

The rule:
$$\irule{\Gamma \vdash T:s} 
           {\Gamma[x:T]~\mbox{well-formed}}
           {s \in \{Prop,Type\}}$$
is replaced by:
$$\irule{\Gamma \vdash T:s} 
           {\Gamma[Qx:T]~\mbox{well-formed}}
           {s \in \{Prop,Type\}}$$
and we add the rule:
$$\irule{\Gamma \vdash a:T~~\Gamma \vdash b:T} 
           {\Gamma[a = b]~\mbox{well-formed}}
           {}$$
Then we extend the system by replacing the rule:
$$\irule
        {\Gamma \vdash T:s~~\Gamma \vdash T':s~~\Gamma \vdash t:T~~T \equiv T'}
        {\Gamma \vdash t:T'}
        {s \in \{Prop,Type\}}$$
by:
$$\irule{\Gamma \vdash T:s~~\Gamma \vdash T':s~~\Gamma \vdash t:T
             ~~T \equiv_{\Gamma} T'}
            {\Gamma \vdash t:T'}
            {s \in \{Prop,Type\}}$$
This defines two new judgements: $\Gamma$ is {\it well-formed using the
constraints} and {\it $t$ has type $T$ in $\Gamma$
using the constraints}.

{\sbd Remark:} A term may be well-typed in $\Gamma$ using the constraints and
still be not normalizable.

{\sbd Definition:} Well-typed Without Using the Constraints

Let $\Gamma$ be a context and $t$ and $T$ be two terms. The term $t$ is said 
to be {\it of type $T$ in $\Gamma$ without using the constraints} 
if there exists $\Delta$ subcontext of $\Gamma$ (i.e. obtained by removing
some items of $\Gamma$) such that $\Delta$ has no
constraints, is a well-formed context and $\Delta \vdash t:T$.

{\sbd Proposition:} If a term is well-typed in a context without using the
constraints then it is strongly normalizable.

{\sbd Definition:} Normal Form of a Context

Let $\Gamma$ be a context, the {\it normal form} of
$\Gamma$ is obtained by putting all the types of variables that are well 
typed without the constraints and all the constraints which terms are 
well-typed without the constraints in $\beta$-normal $\eta$-long form.

{\sbd Definition:} Success and Failure Contexts

A normal context $\Gamma$ is said to be 
{\it a success context} if it has only universal variables, and constraints 
relating identical terms. It is said to be {\it a failure context} if there is 
a constraint relating two normal well-typed ground terms which are not 
identical.

\subsection{An Extended Constructive Engine}

We consider an extension of the constructive engine \cite{Huet89}.
This machine has a register that contains a term and a context which is a 
quantified constrained context. The context is separated in two parts by an 
index:
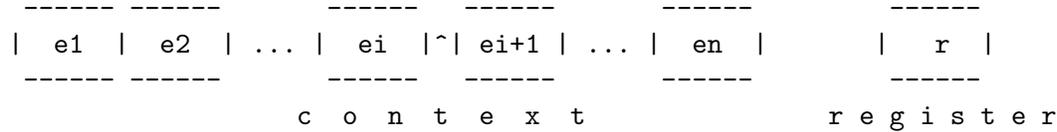
\begin{verbatim}
 ------ ------       ------   ------       ------         ------
|  e1  |  e2  | ... |  ei  |^| ei+1 | ... |  en  |       |   r  |
 ------ ------       ------   ------       ------         ------
                   c  o  n  t  e  x  t                r e g i s t e r
\end{verbatim}
Two invariants are maintained:
the context $[e_{1}; ...; e_{i}; e_{i+1}; ...; e_{n}]$ is a well-formed 
context and the term $r$ is well-typed in the context $[e_{1}; ...; e_{i}]$.

The basic instructions of the machine are:

$\bullet$ move the index on the left or to the right and erase the register,

$\bullet$ check that a term $t$ is well typed in the context 
$[e_{1}; ...; e_{i}]$ and put it in the register.

$\bullet$ check that the type of the term in the register is $Prop$ or $Type$
and insert a declaration of a new variable of this type between $e_{i}$ and
$e_{i+1}$.

$\bullet$ check that $e_{i}$ is the declaration of an existential variable $y$,
erase it, insert a constraint $T = U$ where $T$ is the type of this existential
variable and $U$ the type of the term $r$, insert a constant declaration
$y:=r$, put the context in normal form and remove the constraints relating 
equal terms and fail if the context is a failure context.

In a more sophisticated engine the constraints that are first order
unification problems are automatically solved.
This can be done in simplifying the constraints \cite{Huet75} \cite{Huet76}
\cite{complete} and solving the trivial 
equations $x = t$ where $x$ is an existential variable\footnote{Also
the argument-restricted unification problems 
\cite{Miller91a} \cite{Miller91b} \cite{Pfenning91} can be solved, if we 
simplify equations and solve the trivial equations
$(x~c_{1}~...~c_{n}) = t$ where $x$ is an existential variable and 
$c_{1}, ..., c_{n}$ are atomic terms which heads are distinct universal 
variables declared on the right of $x$.}.

\subsection{Proof Checking with this Engine}

When we verify a development as:
\begin{verbatim}
Theorem I.
Statement (P:Prop)(P -> P).
Proof [P:Prop][x:P]x.
\end{verbatim}
we construct the term $(P:Prop)(P \ra P)$ we declare an existential variable 
$I$ of this type, we construct the term $[P:Prop][x:P]x$ and we instanciate the
variable $y$ with this term, the constraint 
$(P:Prop)(P \ra P) = (P:Prop)(P \ra P)$ is generated and removed since it 
relates equal terms (if the term had not the good type, we would get a
failure context) and a constant $I := [P:Prop][x:P]x:(P:Prop)(P \ra P)$ 
is added to the context.

\subsection{Proof Construction by Tactics with this Engine}

\subsubsection{Application}

The main idea in this section is that applying a tactic is applying a 
substitution to the current context.
For instance let us consider the context:\\
$\Delta = [\fa T:Prop; \fa R:T \ra T \ra Prop; \fa Eq:T \ra T \ra Prop;$\\
$\fa Antisym:(x:T)(y:T)(R~x~y) \ra (R~y~x) \ra (Eq~x~y);
\fa a:T; \fa b:T; \fa ax1:(R~a~b); \fa ax2:(R~b~a)]$
and $\Gamma = \Delta [\ex x:(Eq~a~b)]$.
We declare new existential variables $h_{1}:T$, $h_{2}:T$, 
$h_{3}:(R~h_{1}~h_{2})$, $h_{4}:(R~h_{2}~h_{1})$.

We perform the substitution:
$$x \la (Antisym~h_{1}~h_{2}~h_{3}~h_{4})$$
We get the context:\\
$\Delta  [\ex h_{1}:T; \ex h_{2}:T; \ex h_{3}:(R~h_{1}~h_{2}); 
\ex h_{4}:(R~h_{2}~h_{1});(Eq~h_{1}~h_{2}) = (Eq~a~b)]$\\
The constraint is a first order unification problem, it can be solved
automatically, i.e. the following substitutions can be performed:
$$h_{1} \la a$$
$$h_{2} \la b$$
We get the context $\Delta [\ex h_{3}:(R~a~b); \ex h_{4}:(R~b~a)]$.
We have this way performed a $Apply$ step.

If the constraint on $h_{1}$ and $h_{2}$ were more complicated we would have
proposed substitutions for these variables, so the tactic $Apply$ is not 
limited by a unification algorithm.

\subsubsection{Introduction}

A problem with this engine is that the introduction tactic is not the mere 
application of a substitution to a context, because we have to take into 
account that when we instanciate $h$ by $[x:T]k$ the variable $k$ may be 
substituted by a term where $x$ occurs. 
In the second part of this note we generalize the engine in order to be able 
to use this tactic.

\subsubsection{Explicit Dependencies}

Let us consider a goal $(x_{1}:P_{1}) ... (x_{n}:P_{n})P$

When we apply $n$ times the tactic $Intro$ and then once the tactic $Apply$ 
with a head $u$ we get the incomplete proof
$[x_{1}:P_{1}] ... [x_{n}:P_{n}](u~h_{1}~...~h_{p})$ where the variables
$h_{1}, ..., h_{p}$ must be instanciated by terms where the variables
$x_{1}, ..., x_{n}$ may occur. We may mark explicitly this dependence in
anti-skolemizing these variables and considering the term:
$$[x_{1}:P_{1}] ... [x_{n}:P_{n}]
(u~(h_{1}~x_{1}...x_{n})~...~(h_{p}~x_{1}~...~x_{n}))$$
now the variables $h_{1}, ..., h_{p}$ must be instanciated by terms well formed
in the same context as the initial goal.

This anti-skolemization has to be done after $n$ introductions and one
application. It cannot be done after one introduction because it would give
the substitution $h \la [x:T](k~x)$ i.e. modulo $\eta$-conversion to the
substitution $h \la k$. 

Proof synthesis methods using this elementary substitution (and also more
general substitutions) are studied in \cite{complete}.

This elementary substitution is not very usable for interactive proof 
construction since subgoals appear with unuseful and confusing
dependencies.
We are going now to generalize the engine to permit an introduction tactic.

\section{Introduction: An Engine with Sections}

\subsection{Sections and Discharge Operation}

Sections have been introduced in the framework of proof verification
\cite{deBruijn87} \cite{deBruijn89} (see also \cite{namingscoping})
to simplify the way proofs are written.
Instead of writing:
\begin{verbatim}
Theorem I.
Statement (P:Prop)(P -> P).
Proof [P:Prop][x:P]x.
\end{verbatim}
We write:
\begin{verbatim}
Theorem I.
   Variable P:Prop.
   Hypothesis x:P.
Statement P.
Proof x.
\end{verbatim}
or:
\begin{verbatim}
Section I.
   Variable P:Prop.
   Hypothesis x:P.

   Theorem I.
   Statement P.
   Proof x.
End I.
\end{verbatim}
When such a development is checked, the variable $P$ and $x$ are first 
declared. Then the theorem $I := x:P$ defined and 
{\it when the section is closed} the variables $P$ and $x$ are erased from 
the context and discharged in the constant $I$ to make 
$I := [P:Prop][x:P]x:(P:Prop)(P \ra P)$.
So the successive values of the context are:\\
$[~]$\\
$[\fa P:Prop]$\\
$[\fa P:Prop;\fa x:P]$\\
$[\fa P:Prop; \fa x:P; I:=x:P]$\\
$[I := [P:Prop][x:P]x:(P:Prop)(P \ra P)]$

Extra information must be added in the context to remind the limits of sections
and the locality or globality of objects.

\subsection{Explicit Sections}

Another way to implement sections is to add two new kind of items in
the context: beginnings and ends of sections.  Checking the same development,
the context would take the values:\\
$[~]$\\
$[Begin]$\\
$[Begin;\fa P:Prop]$\\
$[Begin;\fa P:Prop;\fa x:P]$\\
$[Begin;\fa P:Prop;\fa x:P; I:=x:P]$\\
$[Begin;\fa P:Prop;\fa x:P;I:=x:P;End]$

Now when we want to access to the constant $I$ we remark that it 
is inside a closed section where there are two local variables so we discharge 
these variables in the constant and we get 
$I := [P:Prop][x:P]x:(P:Prop)(P \ra P)$.
We cannot access to the variables $P$ and $x$ since they are local to 
a closed section and so they are out of scope.

In the first method the discharge operation is made once when the section is 
closed and access to a variable is simple.
In the second method closing a section is very simple since we just have to
add an item $End$ to the context, but the access to a variable or a constant 
is complicated
since we have to discharge local variable in it at each access.
This can be compared with the call by value or call by name evaluation 
strategy.

\subsection{Introduction Tactic}

In an engine with explicit sections it is possible to have an introduction 
tactic.
Each existential variable is a little section. For instance, let us consider
the context:
$$[Begin;\ex h:(P:Prop)(P \ra P); End]$$
We perform an introduction and get:
$$[Begin;\fa P:Prop; \ex h:P \ra P; End]$$
From the outside of the section, the variable $P$ cannot be seen and $h$ has 
type $(P:Prop)(P \ra P)$ but from the inside of the section
$P$ can be seen and $h$ has type $P \ra P$. 
If we perform another introduction we get:
$$[Begin;\fa P:Prop; \fa x:P; \ex h:P; End]$$
Then an application:
$$[Begin;\fa P:Prop; \fa x:P;  h:= x:P; End]$$
Seen from the outside of the section the value of 
$h$ is $h := [P:Prop][x:P]x:(P:Prop)(P \ra P)$

\subsection{Physical Closing of an Explicit Section}

When a section does not contain any existential variable it can be physically 
closed i.e. the context can be replaced by the equivalent context:
$$[h := [P:Prop][x:P]x:(P:Prop)(P \ra P)]$$

\section{Examples}

A prototype of this system is implemented in an experimental version of the
system Coq. We give here two proofs of a lemma of Tarski's theorem.
The first is written completely top-down. In the second many lemmas are 
given such that the proof of each of them is very short.
\begin{verbatim}
Parameter T:Prop.

Parameter Eq:T->T->Prop.

Parameter R:T->T->Prop.

Axiom Antisym.
Assumes (x:T)(y:T)(R x y) -> (R y x) -> (Eq x y).

Axiom Trans.
Assumes (x:T)(y:T)(z:T)(R x y) -> (R y z) -> (R x z).
 
Parameter f:T->T.

Axiom Incr.
Assumes (x:T)(y:T)(R x y) -> (R (f x) (f y)).

Definition Pre = [x:T](R x (f x)).

Parameter M:T.

Axiom Up. 
Assumes (x:T) (Pre x) -> (R x M).

Axiom Least.
Assumes (y:T)((x:T)(Pre x)->(R x y))->(R M y).
\end{verbatim}

Example 1:

\begin{verbatim}
Theorem Tarski.

Statement (Eq M (f M)).

Apply Antisym.
Apply Up.
Apply Incr.
Apply Least.
Intro. 
Intro. (* x13 *)
Apply Trans.
Apply Incr.
Apply Up.
Apply x13.
Apply x13.
Apply Least.
Intro. 
Intro. (* x27 *)
Apply Trans.
Apply Incr.
Apply Up.
Apply x27.
Apply x27.
\end{verbatim}

Example 2:

\begin{verbatim}
Theorem Tarski.

Statement (Eq M (f M)).

      Remark One.
   
         Variable y:T.

         Hypothesis v.
         Assumes (Pre y).
      
      Statement (R y (f M)).

         Remark Rem.
         Statement (R y M).
         Apply Up. Apply v. 

         Remark Rem'.
         Statement (R (f y) (f M)).
         Apply Incr. Apply Rem. 
  
      Apply Trans. Apply Rem'. Apply v. 

      Remark Two.
      Statement (R M (f M)).
      Apply Least. Assumption One. 

      Remark Three.
      Statement (R (f M) (f (f M))).
      Apply Incr. Apply Two. 

      Remark Four.
      Statement (R (f M) M).
      Apply Up. Apply Three. 

Apply Antisym. Apply Four. Apply Two.
\end{verbatim}


\begin{thebibliography}{99.}

\bibitem{deBruijn87}
N.G. de Bruijn,
The Mathematical Vernacular, A Language For Mathematics With Typed Sets,
{\it Proceedings of the Workshop on Programming Logic}, Marstrand, Sweden,
1987.

\bibitem{deBruijn89}
N.G. de Bruijn,
The Mathematical Vernacular: Examples,
Unpublished manuscript.

\bibitem{Coquand85}
Th. Coquand,
Une Th\'{e}orie des Constructions,
{\it Th\`{e}se de troisi\`{e}me cycle}, Universit\'{e} Paris VII, 1985.

\bibitem{CoqHue88}
T. Coquand, G. Huet,
The Calculus of Constructions,
{\it Information and Computation}, 76, 1988, pp. 95-120.

\bibitem{namingscoping}
G. Dowek,
Naming and Scoping in a Mathematical Vernacular, 
{\it Rapport de Recherche 1283}, INRIA, 1990.

\bibitem{complete}
G. Dowek,
A Complete Proof Synthesis Method for Type Systems of the Cube.

\bibitem{Coq}
G. Dowek, A. Felty, G. Huet, H. Herbelin, Ch. Paulin-Mohring, B. Werner,
The System Coq User's Guide, INRIA 1991.

\bibitem{GoMiWa}
M.J. Gordon, A.J. Milner, C.P. Wadsworth, 
{\it Edinburgh LCF}, Lecture Notes in Computer Science 78, 
Springer-Verlag, 1979.

\bibitem{Huet75}
G. Huet,
A Unification Algorithm for Typed $\lambda$-calculus,
{\it Theoretical Computer Science}, 1, 1975, pp. 27-57.

\bibitem{Huet76}
G. Huet,
R\'{e}solution d'\'{E}quations dans les Langages d'Ordre 1,2, ...,
$\omega$,
{\it Th\`{e}se de Doctorat d'\'{E}tat}, Universit\'{e} de Paris VII, 1976.

\bibitem{Huet89}
G. Huet,
The Constructive Engine, {\it A Perspective in Theoretical Computer 
Science}, Commemorative  Volume for Gift Siromoney, R. Narasimhan (Ed.),
World Scientific Publishing, 1989.

\bibitem{Miller91a}
D. A. Miller,
Unification Under a Mixed Prefix, To appear in {\it Journal of Symbolic
Computation}.

\bibitem{Miller91b}
D. A. Miller,
A Logic Programming Language with Lambda-Abstraction, Function Variables, and
Simple Unification
{\it Extension of Logic Programming}, P. Schroeder-Heister (Ed.),
Lecture Notes in Computer Science 475, Springer-Verlag, 1991, pp. 253-281.
Also  Report ECS-LFCS-91-159, University of Edinburgh, 1991. 
Also to appear in  Journal of Logic and Computation.

\bibitem{Pfenning91}
F. Pfenning,
Unification and anti-Unification in the Calculus of Constructions,
To appear in {\it Proceedings of Logic in Computer Science}, 1991.

\end{thebibliography}
\end{document}